\documentclass[aps,superscriptaddress]{revtex4}
\usepackage{graphicx}
\def\wn{cm$^{-1}$}
\def\I{\overline{1}}
\def\k{{\bf k}}                 
\def\q{{\bf q}}                 
\def\HH{{$\textbf{H}$}}		
\def\CC{{$\textbf{C}$}}		

\begin{document}

\title{POLYMORPHISM,
PHONON DYNAMICS AND CARRIER-PHONON COUPLING IN PENTACENE}

\author{Raffaele G. \surname{Della Valle}}
\author{Aldo Brillante}
\author{Luca Farina}
\author{Elisabetta Venuti}
\affiliation{Dip. Chimica Fisica ed Inorganica and INSTM-UdR Bologna, viale Risorgimento 4,
Bologna University, I-40136 Bologna, Italy}
\author{Matteo Masino}
\author{Alberto Girlando}
\email{girlando@unipr.it}
\affiliation{Dip. Chimica G.I.A.F. and INSTM-UdR Parma, 
Parco Area delle Scienze 17/A, Parma University,
I-43100 Parma, Italy}

\begin{abstract}
The crystal structure and phonon dynamics of pentacene is computed with
the Quasi Harmonic Lattice Dynamics (QHLD) method, based on atom-atom
potential. We show that two crystalline phases of pentacene exist,
rather similar in thermodynamic stability and in molecular density. The
two phases can be easily distinguished by Raman spectroscopy in the
10-100 cm-1  spectral region. We have not found any temperature induced
phase transition, whereas a sluggish phase change to the denser phase
is induced by pressure. The bandwidths of the two phases are slightly
different. The charge carrier coupling to low-frequency phonons
is calculated.    
\end{abstract}

\date{\today}
\maketitle

\vskip1.0truecm

\section{Introduction}
Despite recent progresses \cite{IBM}, the development of electronic devices
based on organic semiconductors remains an open challenge. Considerable
attention has been recently devoted to pentacene, due to the claim of
mobilities  comparable to crystalline silicon, and other extraordinary
properties \cite{schon01}. Such claims have been show to be invalidated
by scientific misconduct \cite{report},  yet pentacene  still holds  the
record  of highest room temperature mobility among  materials for
organic thin-film transistors.  This record mobility of about 1.5
cm$^2$V$^{-1}$s$^{-1}$ has been reached by continuous improvements
in the sample chemical purification \cite{IBM}.  On the other hand,
the crystal and thin film growth also have paramount importance for
device performance, together with a  proper understanding of the motion
of charge carriers inside the crystal. By combining computational
and experimental approaches, we  have addressed the above  problems
for the specific case of  pentacene.  Here we review the main results
reached so far.

\section{Crystal structures}
The polymorphism of pentacene, both in bulk and in  thin film, has been
object of intense experimental \cite{campbell,holmes,mattheus03} and
theoretical \cite{mattiago,venuti02} studies.  We have analyzed the
five complete X-ray structures of pentacene single crystal crystal 
reported so far by computing the "inherent structures",  defined as
the local minima of the potential energy hypersurface \cite{masino02}.
Each structure was modeled  starting from its experimental molecular
arrangements and a common  {\it ab-initio} molecular geometry.
The identity of molecular geometries in all the configurations is a
necessary prerequisite for allowing identical phases to map into
identical potential minima.  The theoretical structures of mechanical
equilibrium were determined by minimizing the potential energy, $\Phi$.
The unit cell axes, angles, molecular orientations  and positions were
freely varied in the minimization. At this stage, we adopted the
rigid molecule approximation, retaining the molecular geometry fixed.
The intermolecular potential was represented by an atom-atom Buckingam
model, with Williams parameters set IV, plus electrostatic interactions
described by the {\it ab-initio} charges residing on the atoms \cite{masino02}.

\begin{figure}
\includegraphics* [scale=0.45]{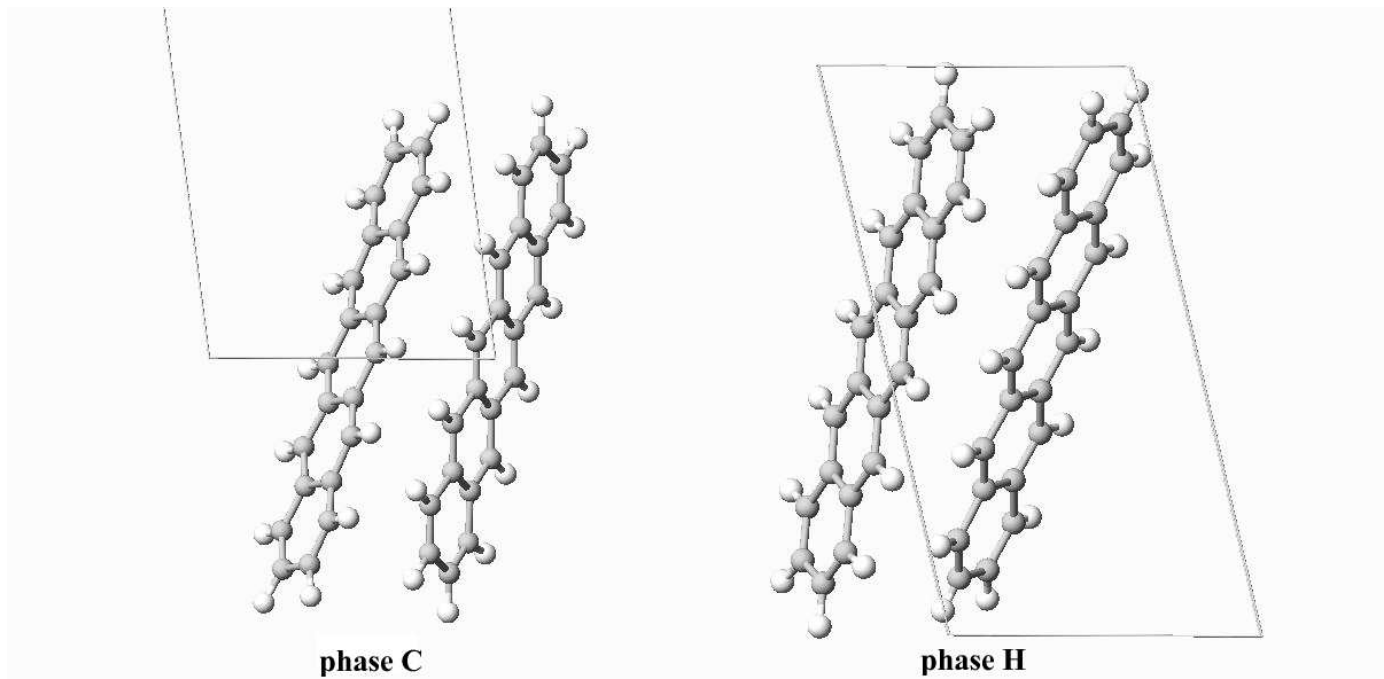}
\caption{Comparison of pentacene {\CC}  and  {\HH}  polymorphs.}
\label{crystal}
\end{figure}

With the above described method, we have been able to show that the five
reported crystal structures correspond to two different ``inherent
structures'', i.e.,  two local $\Phi$ minima.  One of  the  minima, 
labeled as phase {\CC}, corresponds to the structure determined
by Campbell {\it et al.} \cite{campbell}. The other polymorph,
phase {\HH}, corresponds to the structure found in the more recent
measurements \cite{holmes}. Both polymorphs belong to the triclinic
system, space group $P\I$ ($C_i^1$) with two molecules per unit cell
arranged in the usual herringbone pattern. Fig. \ref{crystal} compares the two
structures, which appear to be very similar, the main difference being
a shift of the two molecules along the long molecular axis, yielding 
different  $d(001)$-spacings  (14.5 vs 14.1 Å  for {\CC} and
{\HH} phases, respectively \cite{mattheus03}).

We have extended the analysis by systematically sampling the potential
hypersurface  to gain  information on the global stability of the minima.
In the case of pentacene, a quasi-Monte Carlo search yielded several
hundred of distinct minima, with surprising  variety of structural
arrangements \cite{valle03}. On the other hand, deep minima
are easily accessible since their attraction basin tends to be wide.
We found that the two deepest minima indeed correspond to the
experimental  {\CC} and {\HH}  polymorphs, which are described correctly.
Other deep minima with layered structures, which might correspond
to the thin film polymorphs found to grow on substrates, were also found.

After the above analysis of the potential energy $\Phi$, we have
accounted for the effects of temperature $T$ and pressure $p$ by
calculating the structures of minimum Gibbs energy $G(p,T)$ with
quasi harmonic lattice dynamics (QHLD) methods \cite{masino02}.
In this method the vibrational contribution to the Gibbs energy
is estimated in the harmonic approximation:

$$G(p,T) = \Phi + pV + \sum_{{\q},{\rm j}} \frac {\hbar \omega_{{\q},{\rm j}}}
{2} + k_{\rm B}T \sum_{{\q},{\rm j}} 
\ln \left [1 - \exp \left (- \frac {\hbar 
\omega_{{\q},{\rm j}}} {k_{\rm B}T}\right)\right ] \eqno(1) $$                                                

\vskip0.5truecm
\noindent
Here V is the  molar volume, and  $\omega_{{\q},{\rm j}}$   are the harmonic
phonon frequencies, calculated by diagonalizing the dynamical matrix obtained
from the second derivatives of $\Phi$ with respect to the displacements of
the j-th molecular coordinate with wavevector {\q}.
The second term in Eq. (1)  is the zero-point energy, and the last term is the
entropic contribution. The sums are extended to all phonon frequencies. Since 
pentacene exhibits  {\it ab-initio} intra-molecular
frequencies as low as 38 {\wn}, we had to allow for the coupling between lattice
and intra-molecular vibrations, found to be necessary in similar cases. 
We adopted an exciton-like model \cite{girlando00}, where the interaction between
different molecular coordinates is mediated by the intermolecular potential which,
being a function of the interatomic distance, depends directly on the atomic 
displacements. Since these correspond to the Cartesian eigenvectors of the normal
modes of the isolated molecule, we used the  {\it ab-initio}  eigenvectors and the
scaled {\it ab-initio} frequencies. Intra-molecular modes above 300 {\wn} were not taken
into account, because  coupling is important only for low frequency modes.

Comparison between minimum $\Phi$ and minimum $G$ structures shows that the latter,
which include vibrational effects, brings the calculated volumes from about 4\% to about
2\% from experimental values, or better. Moreover, the minimum $G$ calculations
reproduce correctly the thermal expansion of the {\HH} phase. The experimental volume
expansion from  90  to  293 K  is  3.5 \%, to be compared with a calculated value of 2.7.
The coupling between lattice and intra-molecular vibrations is important, since 
the calculated expansion falls down to 1.9 \% by neglecting it \cite{masino02}.

For both phases, the QHLD calculations exhibit a mechanical instability at $T >$ 550  K,
where no minimum of  $G(p,T)$  can be found and the crystal structure diverges. 
Experimental melting temperatures are given between 540 and 580 K. As discussed
elsewhere \cite{valle98}, loss of mechanical stability in QHLD calculations is
not necessarily coincident with melting, but is often close to it. A good agreement
is also found for the experimental sublimation heat  $\Delta_{sub}H$ = 184 $\pm$ 10 kJ/mol,
which is to be compared with the Gibbs energy calculated at 0 K,  $G_0 \sim$ 174 kJ/mol for both
phases.

\begin{figure}
\includegraphics* [scale=0.65]{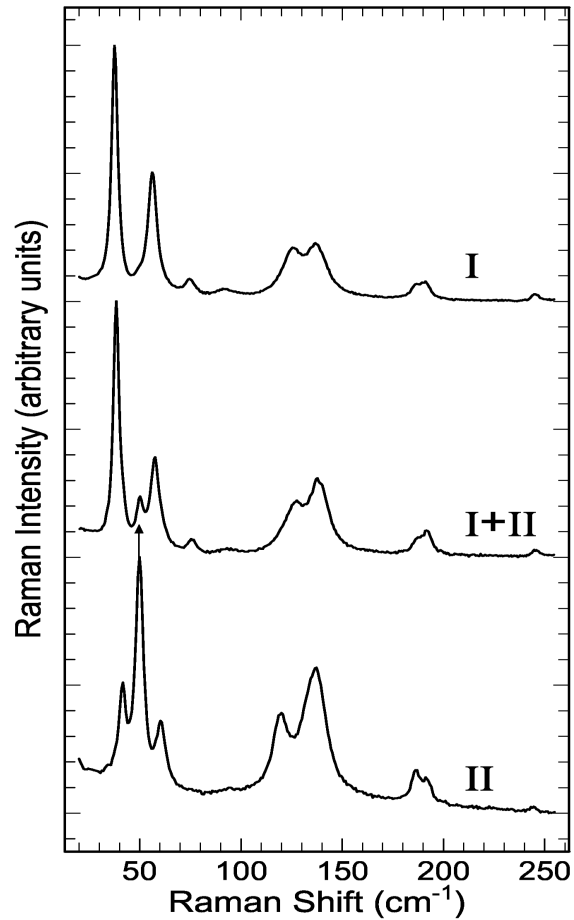}
\caption{Typical Raman spectra of pentacene crystals (from Ref. \cite{brillante02}).}
\label{raman}
\end{figure}

\section{Lattice phonon spectra}

Although the above QHLD calculations  clearly indicate that both {\CC} and  {\HH} polymorphs
are mechanically and thermodynamically stable, one might still harbor
doubts about the existence of the form reported by Campbell  {\it et al.} \cite{campbell},
since all the recent X-ray studies have failed to reproduce it.  Both the minimum $\Phi$ and
minimum $G$  calculations  indicate that the lowest frequency inter-molecular
(lattice) vibrations should differ considerably in {\CC}  and  {\HH} polymorphs.
We have then chosen to use Raman scattering in the lattice phonon region
(10-150  {\wn}).  This choice was also supported by the possibility of interfacing
optical microscopy  to Raman spectroscopy, achieving a spatial resolution of about
1 $\mu$m, which enables a careful mapping of the physical features of each crystal
sample. The spectrometer used was a Jobin-Yvon T64000 triple monochromator,
interfaced to an Olympus BX40 microscope, and with excitation from a Krypton
laser at 752.5 nm. Pentacene crystals (generally metal-like bluish platelets) from
different commercial sources  were used, without further purification, as samples
for the Raman measurements. A different morphology, i.e., tiny blue microcrystals,
was obtained for samples vapor-grown by fast sublimation in Ar or  N$_2$ atmosphere.
The Raman spectra were measured by using the above described different crystal
samples. The crystals 'physical  purity' was checked by mapping their phonon
spectra in several spots of  the sample.

Fig. \ref{raman}  reports typical  Raman  profiles
in the region 20-250 {\wn}, where all the lattice, as well as a number of low-frequency
intra-molecular modes, are present. Indeed, as we have already pointed up,  some
intra-molecular modes occurring in this spectral region  are  coupled  with
lattice  phonons. A  detailed assignment of the spectra  is reported 
elsewhere \cite{brillante02,valleb03}.   Here, we focus on the lowest frequency
bands  (below  100  {\wn})  which differ significantly in spectrum I and II.
In spectrum I, one identifies four bands almost uniformly spread in the considered
frequency range. In the case of spectrum II, instead, the three lowest bands are
clustered in a narrow frequency range around 55 {\wn}, and one very weak band is
found near 100 {\wn}.  This is precisely the frequency pattern we calculated for
polymorph  {\HH} and {\CC}, respectively.  Thus commercial samples, which give
spectrum I, all belong to the {\HH} phase, whereas the samples obtained by fast
sublimation in inert atmosphere give spectrum II and correspond to phase {\CC}. 

A further remark should be made on Fig. \ref{raman}: the spectra originating from commercial
samples (polymorph {\HH}, morphologically platelets) evidence in some cases an
additional feature at $\sim$ 50 {\wn}, as shown by the arrow in the spectrum
(I + II) of the same figure. This band is clearly due to a physical residual
impurity of polymorph  {\CC}  (morphologically microcrystals), that can be monitored
in regions of the commercial specimens extended only few microns, as indicated by
the decreasing intensity of the band on increasing the crystal area  sampled.
The fact that a residual  presence  of  polymorph {\CC} can be found in different
amounts on the measured crystals is certainly related to the different degree
of physical purity of  the used commercial  products.

The Raman indications have been subsequently confirmed by direct X-ray measurements \cite{farina03},
which indeed yield two distinct structures, matching the published measurements for
forms {\CC} and {\HH}.  However,  the Raman tool  is  more  versatile, and allows one to
analyze the phase homogeneity, which, as discussed below, might have important consequences
on the measured mobilities.  Although identified only in recent years, 
the phase {\HH} is the most easily encountered, irrespectively on the method of preparation,
either from sublimation or from solution. One therefore expects it is the thermodynamically
stable form, even if our calculations indicate a almost equivalent stability for
the two phases.  We have then decided  to investigate the possible presence of
phase transitions  as  a function temperature and of pressure by using the Raman
microprobe technique. At the same time, we could verify that the  $T,~p$ evolution of the phonon
frequencies is well reproduced by the  computed QHLD frequencies \cite{valle03,farina03}.

No phase transition was detected by changing $T$  from 400  to  4.2 K \cite{valle03}. On the other
hand, we have found that phase {\CC} starts to irreversibly transform to the denser {\HH}
phase by just applying a moderate pressure of about 0.2 Gpa \cite{farina03}.  However, the
phase transition goes to completion only when pressures up to at least 5 GPa  are  applied.
Although increasing pressure is obviously expected to favor the higher density {\HH}
phase, the transition mechanism cannot be explained on the basis of relative densities only.
In fact,  thermal annealing at $p$ = 0.6 GPa does not  help to speed up the structural change
started at 0.2 GPa. Together with the sluggish evolution of the transition at room temperature,
this is an indication that either a high energy barrier must be overcome,
or that the driving force becomes weaker under pressure.

\section{Charge carrier-phonon coupling} 

Despite intensive research, we still lack of a proper description of the charge carriers
motion in organic molecular semiconductors \cite{pope}.  The problem is a difficult one,
since bandwidths, electron-electron interactions, polarization energies, and
electron-phonon coupling are all comparable in magnitude.  In this paper we shall
simply  try to estimate the relative magnitude of some of the involved parameters. 

\begin{figure}[h]
\includegraphics* [scale=0.75]{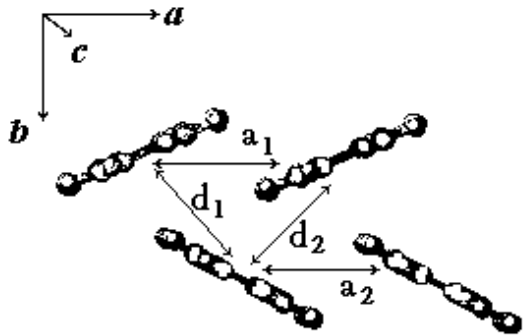}
\caption{Definition of pentacene largest hopping integrals}
\label{hopping}
\vskip0.5truecm
\end{figure}

There is general agreement that the intrinsic mobilities of oligoacenes  exhibit
a hopping-to-band transition around room temperature \cite{bredas03}. Here, we consider
only the bandlike regime, describing the electronic structure in a tight-binding scheme.
We have first calculated the intermolecular hopping integrals between pairs of molecules
as the splitting of HOMO  and LUMO levels from  the isolated molecule to the pair
by using the INDO/S semi-empirical  method \cite{zerner}.  The most important integrals,
labeled  $t_{a_1}$, $t_{a_2}$, $t_{d_1}$, and $t_{d_2}$ , are shown in Fig. \ref{hopping}.

The band dispersions are  obtained by:

$$\epsilon_{\pm}({\bf k})  = (t_{a_1} + t_{a_2}) \cos({\bf ka}) \pm
2 \left\{t_{d_1}\cos[{\k}({\bf a} + {\bf b})/2]
 + t_{d_2} \cos [{\k} ({\bf a} -{\bf b)})/2] \right\} \eqno(2) $$
\vskip0.5truecm

The resulting bandwidths for {\CC} and {\HH} computed structures are reported in Table \ref{bandw}.
The temperature dependence is also given. As it  usually happens in this kind of
calculations \cite{bredas03}, the conduction band (CB)  is wider than the valence
band (VB), contrary to what it would be expected, since in oligoacenes hole mobility
is larger than electron mobility. We believe this is at least in part due to
a computational artifact, since in an Hartree-Fock scheme filled orbital  are better
calculated than virtual orbital energies.   So we concentrate on relative variations.
The lattice contraction by lowering temperature brings about 15 \%  bandwidth increase.
More importantly, a similar difference in bandwidth  exists between polymorph
{\CC} and {\HH}, wider bands being  of course found in the denser  polymorph  {\HH}.
Therefore, {\CC}  micro-domains embedded in the commonly grown {\HH} phase would
likely reduce the band mobility, beside the obvious effects of defect
scattering at the grain boundaries. 

\begin{table}[ht]
\caption{Calculated VB and CB bandwidths as a function of temperature and structure.}
\begin{tabular} {|c|l|ccc|}
\hline
Structure &  &~~ $T$ = 0 K ~~ & ~~ $T$ = 150 K ~~ & ~~ $T$ = 300 K ~~\cr
\hline
  {\CC} &~~ Valence Band (VB)   ~~ &  548 meV  & 511 meV  &  470 meV \cr
        &~~ Conduction Band (CB)~~ &  588 meV  & 548 meV  &  508 meV \cr
\hline
  {\HH} &~~ Valence Band (VB)   ~~ &  632 meV  & 585 meV  &  536 meV \cr
        &~~ Conduction Band (CB) ~~ &  672 meV  & 624 meV  &  576 meV \cr
\hline
\end{tabular}
\label{bandw}
\end{table}
\vskip 0.6 cm

Extensive theoretical and experimental studies have been devoted to the characterization
of  charge carrier coupling to molecular vibrations, which modulate on-site energies
(Holstein coupling) \cite{bredas02}. On the other hand, very  little is know about the coupling
to inter-molecular, or lattice,  phonons  (Peierls coupling), despite the fact that
lattice phonons are thought to affect mobilities even at low temperatures {\cite{pope}.
In molecular crystals like pentacene, lattice phonons are expected  to couple to
charge carriers mainly by modulating the amplitude of inter-molecular hopping integrals.
The corresponding linear Peierls coupling constants are defined by:

$$g({\q},{\rm j}) = \sqrt{\frac{\hbar}{2\omega_{{\q},{\rm j}}}}
\left ( \frac {\partial t}{\partial Q_{{\q},{\rm j}}} \right )_0 \eqno(3) $$ 
\vskip0.5truecm
\noindent
where $ Q_{{\q},{\rm j}}$ is the normal coordinate for the j-th phonon with wavevector
{\q}. We have assumed the low-frequency phonons, mainly inter-molecular in character,
as dispersionless,  and computed the Peierls constants for the {\q} = 0 eigenvectors only.
Coupling to acoustic phonons, which is zero at {\q} = 0, is not considered. Within
these approximations, symmetry arguments show that only the totally symmetric ($A_g$)
phonons can be coupled to charge carriers. To evaluate the Peierls constants we have
calculated the hopping integrals, as detailed above, for the equilibrium and for
geometries displaced along the QHLD eigenvectors. The $g({\q},{\rm j})$ are then
obtained by numerical differentiation.  In Table \ref{g} we report the total Peierls coupling
constants (sums of the modulations of the four hopping integrals defined
in Fig. \ref{hopping})  for the {\HH} phase at 0 K. 

\begin{table}[ht]
\caption{Low-energy $A_g$ phonons  and  Peierls coupling  constants  of  {\HH} pentacene  at 0 K.}
\begin{tabular}{|cccc||cccc|}
\hline
~~$\omega$/{\wn}~~ & $g_{\rm VB}$/meV~~ & ~~$g_{\rm CB}$/meV~~ & & &
~~$\omega$/{\wn}~~ & $g_{\rm VB}$/meV~~ & ~~$g_{\rm CB}$/meV~~ \cr
\hline
  ~27  &   8.2   &  7.2   & & &   162    &  1.5   &    1.6  \cr
  ~61  &   7.1   &  4.9   & & &   203    &  7.2   &    5.5  \cr
  ~70  &   0.5   &  3.1   & & &   206    &  2.8   &    3.0  \cr
  ~96  &   3.7   &  5.5   & & &   238    &  3.7   &    1.5  \cr
  132  &   1.1   &  1.0   & & &   241    &  1.0   &    1.5  \cr
  148  &   6.9   &  0.0   & & &   259    &  5.4   &    2.8  \cr
  156  &   3.8   &  5.9   & & &   261    &  3.8   &    2.4  \cr
\hline
\end{tabular}
\label{g}
\vskip2.truecm
\end{table}

Table \ref{g} shows that several phonons are appreciably coupled to the charge carriers,
and that some difference exists between the coupling in the VB and in the CB. 
The overall coupling strength, as measured by the lattice relaxation energy, 
$E_{\rm LR} = \sum_j g_j^2 / \omega_j$, is 35.3 and 26.9 meV for the VB and the
CB band,  respectively.   The  largest contribution  to $E_{\rm LR}$  ($\sim$ 70 \%)
comes from the two lowest frequencies modes,  which are described as librations  around
the axes perpendicular to the long molecular axis, i.e., approximately parallel to the
conducting  $ab$  plane (Figs. \ref{crystal} and  \ref{hopping}).   Given the low
frequency of these librations, they are likely to affect mobilities even at low temperatures.
A more detailed analysis on the pentacene mobilities and their temperature dependence
is in progress.

\begin{acknowledgments}
Work supported by the Consorzio Interuniversitario Nazionale per la Scienza e
Tecnologia dei Materiali  (I.N.S.T.M.)  under the Prisma 2002 project,  and by the Italian
Ministero dell' Istruzione, dell'Universit\`a e della Ricerca (M.I.U.R.).
We acknowledge enlightening discussions with Karl Syassen.
\end{acknowledgments}

\end{document}